Dr. Matias Slavov
Postdoctoral researcher
Tampere University

# Eternalism and Everettian Quantum Mechanics

*Abstract:* This paper shall explore the conjunction of eternalism and Everettian quantum mechanics. It shall be argued that there is a strong analogy between these two views. In case there is an indefinite number of worlds and observers that are all equally real, there should be an indefinite number of local times which are all also equally real. Whereas Everettianism, specifically the diverging version, treats actuality indexically, relativistic eternalism treats tense indexically. All times exist analogously to all isolated Everettian worlds. There is no unique 'now' that cuts throughout all that physically exists. Instead, as eternalism propounds, all times exist. The paper concludes that eternalism and the many-worlds interpretation are not only compatible but complement each other, providing a coherent framework for understanding the nature of temporal reality.

*Keywords:* metaphysics of time; eternalism; many-worlds interpretation; modal realism

## 1. Introduction

Eternalism predicates the existence of all times: past, present, and future. All things and events, whether past, present, or future, exist unconditionally. This stands in contrast to presentism, which asserts that only the present things exist unconditionally, and to the growing block view, which maintains that the past and the present things exist unconditionally, but not future ones. Under eternalism, the whole four-dimensional block world exists from its beginning to its end. There is a distinction between the B-theoretic account of eternalism and the A-theoretic moving spotlight view. Both take all tensed locations to exist, but the former treats them as indexicals, while the latter treats them as absolutes. This article focuses on eternalism in the B-theoretic sense, as that version of eternalism maintains the perspectival nature of the 'now'.

It is typically held that relativistic physics is consistent with eternalism. At least relativity debunks presentism and the growing block view if they require a completely universal 'now' stretched throughout the entire universe. Such a universal time does not fit with the relativity and the conventionality of simultaneity. Many have extensively argued that both the special and general relativity are in tension with presentism (some 21st century examples on special theory: Saunders 2002; Balashov and Janssen 2003; Peterson and Silberstein 2010; Wüthrich 2010, 2012; Fazekas 2016; Dyke 2021, and on general theory: Romero and Pérez 2014; Read and Qureshi-Hurst 2021; Baron and Le Bihan 2023). There is no reason to rehearse these arguments, as it is generally assumed that relativity is apt for eternalism.[1] It has not been extensively explored, however, whether quantum mechanics agrees with eternalism. Importantly for the metaphysics of time, some considerations from quantum mechanics might lend support to the idea that the future is open and non-existent. This is clearly at odds with eternalism which maintains the existence of all times.

This article centers around Everettian many-worlds quantum mechanics, specifically a metaphysical account based on a Lewisian modal realism. I shall not side with any philosophical interpretation of quantum mechanics. The focus is instead on the concept of time. To that end, the following structure is applied. The next subsection introduces the quantum challenge to eternalism, as some formulations might indicate the non-existence of the future (§2.1). Then the core commitments of many-worlds position are laid out, and the relationship between eternalist account of time and the Everettian theory are considered (§2.2). Alternative, non-eternalist versions of Everettianism shall also be discussed (§2.3). The concluding section claims eternalism is well-suited for Everettianism (§3). This connection remains robust whether we consider the diverging, multiverse version or the overlapping, single-universe version. Different worlds of the multiverse or different branches of the Big-Bang-to-the-end-of-time universe are completely separated and have their own times. The fusion of Everettianism and eternalism suggests that all worlds and all times exist. Therefore, the article concludes that the arguments for eternalist metaphysics are compelling not only within the well-trodden relativistic setting but also in the quantum context.

---

[1] This characterization omits some typical objections. The conventionality of simultaneity in special relativity has been used to challenge the Rietdijk–Putnam argument for eternalism (see, for example, Ben-Yami 2006, Dieks 2012 and Rovelli 2019). There is also the so-called neo-Lorentzian interpretation of relativity (Craig 2001, Zimmerman 2007) that retains absolute simultaneity. Some complications arise due to the so-called Friedman–Lemaître–Robertson–Walker cosmological solution of general relativity (Swinburne 2008), as this enables one to argue for cosmic simultaneity, for a unique hypersurface of cosmic time.

## 2. Quantum mechanics and eternalism

### 2.1 Does quantum mechanics predicate an open and non-existent future?

Putnam's classical argument for the four-dimensionality of the world (1967) suggests that statements about the existence of definite physical events, irrespective of their spacetime location, have truth value 1 or 0. The observer's spacetime location and hence their temporal relation to the event does not matter. The event can be in the observer's past, present or future. Sudbery (2017, 4429) has developed a different argument: "According to quantum mechanics, statements about the future made by sentient beings like us are, in general, neither true nor false; they must satisfy a many-valued logic". Future-tense propositions have truth values ranging between 0 and 1. Much like Aristotle and Prior before him, Sudbery thinks there are genuine future contingents, and the future is indeterminate. There are no future events and so no bivalent truth values concerning statements of those potential events. It should be clarified that this is not the only option in the debate. The idea that indeterminism can seamlessly coexist with the existence of the future and hence eternalism has a well-established history, tracing back to Montague in the 1960s.

Sudbery's argument can be supported by considering the probability density associated with a particle. We cannot know beforehand, not even in principle, where a particle like a photon will be located on the detector screen in the double slit experiment. All we can do is to assign degrees of probability for the detection. The probability of detecting a photon, $Prob(in\ \delta x\ at\ x)$, is directly proportional to the square of the light wave amplitude, $|A(x)|^2 \delta x$. Accordingly, the probability density, $P(x)$, is directly proportional to the square of the light wave amplitude. The probability density for detecting a photon is equally proportional to the square of the amplitude function of the corresponding electromagnetic wave. This proportionality connects the wave and the particle nature of light. To connect the probability density with the wave function, $\psi(x)$, the basic wave function equation is $P(x) = |\psi(x)|^2$. A simple, one-dimensional example is provided by a particle in a box of length $l$, sketched in Figure 1.

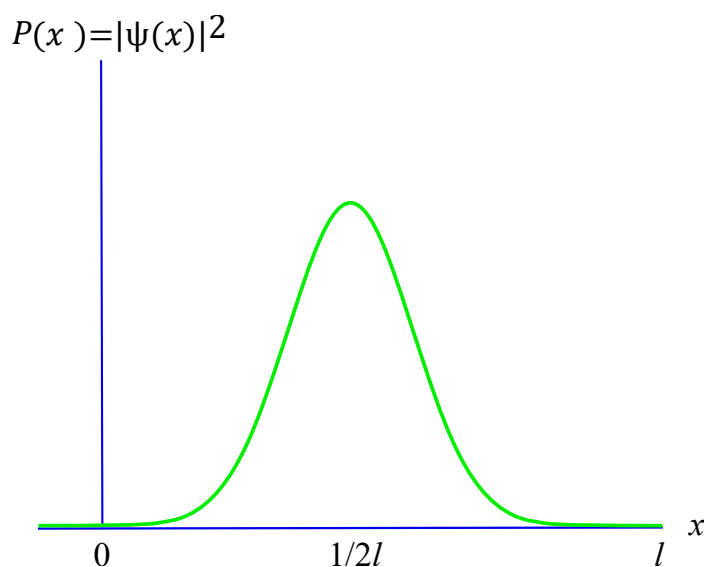

Figure 1. Probability density for a particle in the ground state.

There are different degrees of probability regarding the particle's potential location. The highest probability is at the peak of the curve, at a finite interval approximating $x = 1/2l$. The probability for detecting the particle decreases getting closer to 0 or *l*. The probabilistic interpretation of the wave function equation fits nicely with a presentist or a growing block metaphysics of time. The present and perhaps the past are real, but the future is not. We can assign probabilities, and in some cases successfully predict future outcomes. Probability is not due to our ignorance but due to the nature of time: future is indefinite. Accordingly, there is some probability that a particle will be found somewhere because its future location does not exist in the present.

The indefiniteness of the future, I assume, conforms to our every-day intuitions quite well. Next week's winning lottery numbers do not exist yet. There is only a certain probability that by guessing a series of numbers the machine will randomly produce them. By contrast, the eternalist must hold that the next week's jackpot exists in one definite way. As for the eternalist all events, or all parts of the natural world exists unconditionally,[2] it is only from some perspective they can be denoted to be past, present, or future. The lottery numbers picked out by the machine exist tenselessly; in no sense can they be non-existent and then become existent. If the future is genuinely open and non-existent, as the probabilistic interpretation can be used to suggest, eternalism turns out to be false.

There are many other familiar reasons to think that the past is fixed and cannot be modified, whereas the future is unsettled and open. Our sense of agency, the difference between memory and anticipation, and the determinateness of past causes as opposed to yet non-existent future effects are all hospitable to the idea that the future does not exist. A more fundamental reason for maintaining the non-existence of future could come from stochastic nature of fundamental quantum laws.[3] Under closer scrutiny, the reference to indeterministic laws is in itself rather weak argument for presentism or the growing block view. Indeterminism does not establish an asymmetry between past and future — which is what a probabilistic account of the non-existent future should establish — as the past could be open as well. If indeterministic laws do not entail a unique future, they might still entail a plurality of futures, perhaps different branches of reality, and so indeterminism does not simply rule out the existence of the future.

---

[2] Here I am paraphrasing one of Le Bihan's (2020) atemporal definitions of eternalism.
[3] Glick (2018) argues against quantum indeterminacy.

To argue that the future does not exist, something stronger than familiar every-day intuitions and stochastic laws are needed: objective indeterminism from an objectively present perspective. Mariani and Torrengo (2021) invoke the notion of strong indeterminate present. Although Mariani and Torrengo do not take an explicit stance regarding temporal ontology in their paper, the argument they develop is perhaps the strongest argument for the non-existence of the future. At the moment when the future becomes the present, there is no indeterminism anymore. It is from the present perspective that we can say there is indeterminacy in the future: "the indeterminacy is in the future only until it does not arrive — it is from the point of view of the present, so to say, that the future lacks determination" (Mariani and Torrengo 2021, 3924–7).

To bolster their case about the openness of the future, Mariani and Torrengo consider the spontaneous collapse theory. In the double-slit experiment, if a detector is placed to track the motion of the particle, the interference pattern at the screen will disappear. Perhaps the wave-function collapses. This reading denies the objectivity of the wave-function and the existence of genuine superpositions. Say an electron is in an eigenstate, it has a certain spin, up or down, along the $x$-axis, $e = |\uparrow\rangle$ or $e = |\downarrow\rangle$. The superposition of spin up and down is $e = |\uparrow\rangle + |\downarrow\rangle$. The measurement problem relates to the fact that macroscopic objects seem never to be in a state of superposition. Schrödinger's cat is not alive and dead at the same time. Mariani and Torrengo are sympathetic to von Neumann's collapse postulate: in the action of the measurement, the dynamics of the system breaks down. Macroscopic superposition does not exist. In reference to the GRW-model (Ghirardi, Rimini, and Weber 1986), the wave-function collapses immediately, and the evolution of the system is governed by a new law. That new law is neither linear, nor deterministic, nor subject to time-reversal invariance. The collapse is spontaneous, so all particles have some probabilities per unit time in undergoing a hit on the detector screen. The precise location of the particle hit on the screen is a random fact. The probability distribution is provided by the square of the wavefunction amplitude before the hit (Mariani and Torrengo 2021, 3938). The relevance to the nature of time is clear:

> What is crucial here is that collapse is an intrinsically temporal notion — we cannot make sense of collapse without distinguishing between the time before its occurrence, and the time afterwards. […] The locus of future unsettledness is the intrinsic indeterminacy that certain present states of affairs display. Superposed states are inherently "unstable", and tend to *evolve* into one of the open "options" that inhabit their future (Mariani and Torrengo 2021, 3941–2).

To sum up, this subsection has introduced at least two major challenges to an eternalist account of the nature of time. i) Eternalism maintains the existence of all times: the past, the present, and the future. If the non-existent future comes with a probability for its existence from the viewpoint of the present, eternalism could not be right as it does not ontologically privilege the past or the present over the future. Eternalism, apart from the moving spotlight view, does not include any objective change, that is, a robust passage of time from the future to the past via the present. ii) Eternalism is strongly motivated by the relativity and the conventionality of simultaneity that deny the existence of a privileged hyperplane of simultaneity and absolute present moment. If the future does not exist, but it has a certain probability of occurring based on the present state of the world, the present moment should be universally the same everywhere. Collapse might require a preferred foliation (see Lucas 1999, 10), thus contradicting time dilation which already connects with the relativity of simultaneity.[4]

Next, I consider Everettian quantum mechanics in relation to eternalism. My objective is not to decide which reading of quantum mechanics, or even which version of Everettianism, is right. The objective is rather to consider a suitable quantum match for eternalism. The conjunction of eternalism and the many-worlds interpretation not only complement each other, but they provide a coherent understanding of temporal reality.

**2.2 Everettian quantum mechanics and eternalism**

When quantum experiments are carried out, Everett (1957, footnote to 459–60) originally suggested, different branches of reality emerge with their respective eigenstates. All branches, the elements of a superposition, are equally actual. One branch is as real as any other. No observer will ever be aware of a branching process. The wave function does not collapse. All quantum measurements correspond to multiple measurement results that genuinely occur in different branches of reality.

The relation between relativistic spacetime theories and Everettian quantum mechanics have been discussed previously in the literature.[5] Under orthodox reading of relativity, there is no preferred frame of reference or a privileged foliation of spacetime. Wallace (2001, 16) argues for the following analogy. Instants of time are less fundamental than spacetime, and the branching worlds are less fundamental than the universal state. One does not have to lean on any preferred foliation/basis.[6] There are different times and different worlds. They are all equally real.[7]

---

[4] Provided in some frame two events are simultaneous, $\Delta t' = 0$, in which $\Delta t$ denotes the time-difference of the two events. For another frame, the same events are successive, $\Delta t \neq 0$, because $\Delta t = v(\Delta x')/c^2\sqrt{1-(\beta^2)}$, $\Delta x' \neq 0$, $v \neq 0$. Generally speaking, if two events are simultaneous in one frame, they are successive in a second frame, and they might be successive in different order in a third frame.

[5] For example by Saunders (1998) and Wallace (2001). Saunders (1996) examines the applicability of tense to quantum mechanics more specifically, and Butterfield (2013, 224–5) mentions the advantages of Everettianism regarding time.

Wilson (2020) has advanced a many-worlds interpretation that hinges on quantum modal realism. He supports the diverging multiverse view in opposition to the overlapping, single branching/splitting universe view. The basic metaphysical framework comes from Lewis. This is surprising at first, since Lewis could be interpreted as making somewhat derogatory remarks on quantum physics. It is useful to add i, ii, and iii below to analyze his critical stance:

> I am not ready to take lessons in ontology from quantum physics as it now is. First I must see how it looks when it is purified of instrumentalist frivolity, and dares to say something not just about pointer readings but about the constitution of the world (i); and when it is purified of doublethinking deviant logic (ii); and most of all when it is purified of supernatural tales about the power of the observant mind to make things jump (iii) (Lewis 1986a: ix).

In the view of Lewis, quantum mechanics has three shortcomings that render it superficial for metaphysics: i) it is affiliated with an instrumentalist philosophy of science which is concerned with measurement device readings, that is, with mere predictions instead of understanding the deep structure of reality; ii) its formalism is read through paraconsistent logic; iii) it is involved with downward causation according to which conscious observers affect the motions of elementary particles. Everettian reading of quantum physical experiments is not committed to any of the points i, ii, and iii. It is a realist theory that takes macroscopic superposition literally (in opposition to point i). It does not lean on any kind of dialetheism (in opposition to point ii). It does not elevate the human mind above the physical (in opposition to point iii). Everettianism is, according to Wilson, a good scientific starting point for Lewisian modal realism.

---

[6] Wallace (2001, 22) clarifies that in "everyday circumstances, there exists an (approximate) natural choice of choice of spacetime foliation", as well as "an (approximate) natural choice of basis" but the details of the foliation/basis "are arbitrary, both when examined closely and with respect to spatially remote areas."

[7] Wallace (2001, 23) thinks that we are more familiar, or have been familiar for a longer time, with Minkowski spacetime than with the universal state. He speculates that for this reason the idea of many-worlds strikes more peculiar than the idea of multiple existing times. One might nevertheless say that the Everettian view is less weird than some of the early interpretations, like the Copenhagen interpretation, because it denounces the putative sharp dividing line between the macroscopic and microscopic realms. To quote from Carroll (2021: 231): "when we explicitly move to quantum mechanics, physicists generally start by taking a classical theory and quantizing it. But nature doesn't do that. Nature simply *is* quantum from the start; classical physics, as Everett insisted, is an approximation that is useful in the right circumstances."

Wilson considers the causal-epistemic challenge as anticipated by Skyrms (1980). In what way can we know about other modally real worlds if we do not causally interact with them? How can Lewisian modal realist respond to the epistemic challenge? The atomic theory is corroborated, say, by Brownian motion, as ultimately we humans and our experimental technology are also made of atoms and are connected with the experimental results (Wilson 2020: 10–11). Everett worlds, according to the diverging model, are completely separate and causally isolated. Causal sequences take place only within individual worlds, they do not cut throughout different worlds. Travel between the different worlds is in principle forbidden.

As we do not have access to different worlds, Everettian quantum mechanics cannot be corroborated based on causal interaction, like manipulation of the atomic structure of matter. Instead of causal interaction, the proponent of many-worlds may refer to consistency. Wilson (2020, 60) applies a jigsaw puzzle metaphor. Pieces are shaped the way they are because they fit with other pieces next to them. Adjacent pieces do not cause a particular piece to be how it is, yet there is a non-causal explanation of the shape of the piece based on its neighboring pieces. The features of Everett worlds are non-causally explained by worlds that are close apart. Everett worlds are thus parallel universes that make up one gigantic jigsaw. This fills up possibility space with every different multiple histories permitted by physics. Each world has its own four-dimensional spacetime. From the viewpoint of its inhabitants, their own world is governed by indeterministic laws. One of these four-dimensional spacetimes is ours, while other worlds are analogous but not identical to it (Wilson 2020, 23).

Importantly for this paper, there is a similarity between indexicality of tense in relativistic eternalism and indexicality of actuality in diverging Everettian quantum mechanics. In Wilson's definition, all worlds are actual according to its own inhabitants, and only according to them. The sum totality of these worlds makes up the Everettian multiverse. Every Everett world is qualitatively discernible. Indeterministic quantum phenomena turn out in various ways in different worlds. All possible quantum mechanical sequence of events occurs (Wilson 2020, 22–3). The thesis of indexical actuality succinctly combines Everettian quantum mechanics and Lewisian modal realism:

> The actual world is not intrinsically metaphysically privileged according to modal realists: it is simply the Everett world that you and I happen to inhabit. Everettians recognize no sense in which any Everett world is physically or metaphysically privileged over any other, so correspondingly in quantum modal realism there is no sense in which any metaphysically possible world is privileged over any other. In this respect, quantum modal realism directly mirrors Lewisian modal realism (Wilson 2020: 68).

We can take sentences from the quote above and replace them almost verbatim with a characterization of eternalist metaphysics: *Eternalists recognize no sense in which any local moment 'now' is physically or metaphysically privileged over any other*. The eternalists treat tenses as indexicals (with the exception of the moving spotlight view, which treats tenses as absolutes). Physical events are primitive, but their A-properties are perspectival. When Mary Shelley was writing the *Frankenstein*, for her the wars of the twentieth century will be in the future. For the soldiers crawling and shooting in the battlefield, the Great War is happening right now. For the reader of this article the First World War is in the past. There are definite tenseless events like massive wars, and their reality value is unique, it is not an open interval between 0 and 1.[8] Depending on the perspective — according to whom, where and when — the events are either past, present, or future.

The indexicality of tense can be clarified by perspectival realism about the 'now' (Slavov 2020). It is quite evident, already from common sensical point of view, that spatial location 'here' is an indexical notion. Whether 'now' is an indexical notion is more controversial. If we accept the lessons from relativity, the local moment 'now' cannot be extended everywhere in the universe, at least not in any unique way. It is not that objects or events are automatically affiliated with some tense. My work desk does not have a property of being 'here' or being 'now'. Instead, these spatial and temporal notions are perspectives the observer has to this particular object. Compare this to relativistic Doppler effect. The specific spectrum frequency is, due to relative motion and gravitation, perspectival. In my current frame, my work desk is yellowish. Whether it is red (receding observer) or blue (approaching observer) is a matter of designating the frame of reference. 'Red', 'yellow', or 'blue' are not properties of this object. 'Left', 'right' or 'front' are not properties of the desk. Neither are 'past', 'present' or 'future'. Spatial locations and spectrum frequencies are not absolute, or any way privileged within the eternalist-relativistic account of the world. Neither are temporal locations.

Note that the basic point about indexicality of tense should hold even in the absence of spacetime. If spacetime breaks at some high-energy scale and is hence effective (Crowther 2016, 13), we are not licensed to say that, in any fundamental way, events are contained in spacetime (Knox 2013, 347). This challenges some definition of eternalism, according to which events are tenseless entities spread throughout spacetime, and the observer's contingent spacetime location determines the specific tense. Le Bihan (2020) has developed a metaphysics he calls atemporal eternalism. Even in the absence of spacetime, one can say that all the parts of the natural world exist unconditionally. If deep-down there is no spacetime, then certainly there is no privileged simultaneity slice required by presentism or the growing block view. Observers in different regions of the world have their own times. They may very well disagree as whether something has happened, is happening, or will happen. There is nothing contradictory about this when tense is understood indexically. All tensed locations are equally real. This is, in a nutshell, what eternalism is about: it predicates the existence of all times.

---

[8] Accordingly, the truth value of propositions concerning the existence of such events is uniquely 1, according to eternalism.

Here we have a relevant match between eternalism and Everettianism. There is no 'now' which is the same everywhere in all of physical reality, but different worlds/branches have their own times. Different worlds within the Everett multiverse or different branches within the single universe are causally isolated. There is a clear analogy to relativistic spacelike separation: different locations in the universe are in each other's absolute elsewhere, not connected by any privileged hyperplane of simultaneity. There is no unique present moment that cuts throughout everything that exists and which defines all that exists at that instant. Instead, as eternalism suggests, all times exist.

There is one subtlety within Everettianism that should be addressed, specifically in relation to indexicality. The overlapping interpretation maintains that the universe has, at earlier times, common segments which subsequently split. The diverging view is committed to the existence of parallel worlds; instead of a single universe, there is a multiverse. With the two versions of the theory, there is a difference between what world is an actual world. For the diverging view, observers in the parallel worlds call their own worlds the actual world. Under the overlapping view, actuality is not indexical or relative because there is only one world with all its branches that is actual. If the overlapping view would be judged a stronger version of the Everettian theory, the analogy between the indexicality of actuality and the indexicality of the present could break down.

Whether we are talking about branching/splitting worlds as in the case of the overlapping view or about the multiverse as in the case of the diverging view, the basic point about the indexicality of tense is not jeopardized. Let's briefly consider the fission version of Everettianism, arguably something Everett himself had in mind in the footnote referred to previously and developed further by Greaves (2004) and Tappenden (2008). There is just one Big-Bang-to-the-end-of-time universe. Perhaps Everett's original idea was not about many worlds but about one branching world. There could be just one quantum block universe (the term is from Saunders 1993, 1560). That would nevertheless also have spacelike separated regions with their own times. Everettianism is still consistent with orthodox relativity, it does not introduce a privileged foliation. A quantum block is as eternalist as the Einsteinian block. If we assume that all events in other branches of the single universe exist, what is the difficulty in denying that all times exist? Whether we are dealing with overlapping or diverging Everettianism, they are both consistent with eternalism.

**2.3 Alternative, non-eternalist Everettian approaches**

Conroy (2018) departs from Wilson's modal realism. Instead, her preferred reading of Everettianism relates to Plantinga's actualism. As explained previously, my aim in this paper is not to decide which interpretation of quantum mechanics is right. Likewise, I wish not to decide which reading of Everettianism is correct, yet alone decide what Everett himself thought. Conroy discusses many issues related to the space of possibilities, on how one should understand this issue in relation to Everett's original proposal, including whether Everett was a realist/operationalist considering the wave function, and whether that implies a so-called relative facts interpretation. For this paper, the main focus is on the concept of time.

To quote from Plantinga (1987, 196), "actualism is the view that there neither are nor could have been any entities that do not exist". This general claim certainly does not rule out modal realism in any way. Everything exists, there are no things that do not exist. Conroy connects this to a single, actual universe Everettianism. There are not many worlds, there is only one world. What could have been does not exist; what might become to be does not exist. Although the metaphysics of time is not analyzed in her paper, this kind of position is close to presentism. Past and future events have reality values of 0, only the presently existing things exist. They are the only things with a reality-value.

Plantinga considers possible worlds with an example of a world *W* in which Socrates was not Plato's teacher. If such possible existed, this would

> suggests that all of the possible worlds (*W* included) are somehow simultaneously "going on"—as if each world were actual, but at a different place or perhaps (as the best science fiction has it) in a "different dimension". (Plantinga 1987, 96)

Both the modal realist and the actualist account agree that there is no epistemic access or causal influence across different worlds. That is in principle forbidden. Plantinga notes that the idea of possible existent worlds

> also suggests that I must be able to look into *W* and sift through its inhabitants until I run across one I recognize as Socrates—otherwise I cannot identify him, and hence do not know whom I am talking about. But here the picture misleads us. For taken literally, of course, this notion makes no sense. There is no such thing as "looking into" another possible world to see what is going on there. There

> is no such thing as inspecting the inhabitants of another possible world with a view to deciding which, if any, is Socrates (Plantinga 1987, 96).

There are at least two deeply problematic assumptions in the above characterizations when applied to eternalist Everettianism. 1) All possible worlds exist, or would have to exist, simultaneously. As we are talking about completely isolated worlds, there is no dimension-cutting unique present that all possible worlds share. That would require absolute simultaneity not only in our universe but in all the possible universes. 2) Plantinga presents a *reductio* type of argument: As the different worlds are completely isolated, one cannot go and look into another world. That is, however, true also in the many-worlds interpretation. Looking into different worlds would require a causal relation: an event happens, and some information is sent from that event, which is then received and processed in a finite time. Special relativity forbids causal interactions among spacelike separated events. Seeing into another universe does not happen. But this *reductio* does not succeed against eternalist Everettianism. The many-worlds interpretation is not absurd. It can explain why observing and gathering information from different worlds is impossible.

Despite other relevant readings of Everettianism and despite different versions within the many world interpetation, the eternalist version of Everettianism can retain the core reasons for embracing eternalism. The denial of universal present and the isolation of all the worlds support the locality of the present time, which is an essential tenet of eternalism.

## 3. Coda

There are some reasons to think that quantum mechanics does not go along with eternalism. The probabilistic interpretation of the truth-values of future-tensed statements and the spontaneous collapse interpretation align with an unsettled, non-existent future. This is not the only relevant interpretation, but it nevertheless lends support to presentism/growing block. For its part, Everettian quantum mechanics accommodates eternalism. The Everett interpretation does not affect the standard arguments for eternalism from relativity as it is a relativistically covariant and deterministic theory. The indexicality of actuality is analogous to the indexicality of tense, and the causal isolation of Everettian worlds in the Everett multiverse, or the causal isolation of different branches of the single Everett universe, is analogous to spacelike separation of events. There is a reasonable interpretation of the measurement problem that maintains the existence of all quantum mechanical non-zero amplitude outcomes much the same way as there is a reasonable metaphysics of time which maintains the existence of all times. Whether we are dealing with a diverging, multiverse account or the overlapping, single-universe branching account of Everettianism, different parts of the multiverse/Big-Bang-to-the-end-of-time universe are completely separated and possess their own times. Maybe all worlds/branches and all times exist.

**References:**

Balashov, Y. and M. Janssen (2003) "Presentism and Relativity." *British Journal for the Philosophy of Science* 54: 327–46.

Baron, S. and B. Le Bihan (2023) "Trouble on the Horizon for Presentism." *Philosophers' Imprint* 23 (2): https://doi.org/10.3998/phimp.823.

Ben-Yami, H. (2006) "Causality and Temporal Order in Special Relativity." *The British Journal for the Philosophy of Science* 57: 459–79.

Butterfield, J. (2013) "On Time in Quantum Physics." In Dyke, H. and Bardon, A. (eds.) *A Companion to Philosophy of Time*, 220–41. Chichester, West Sussex: Blackwell.

Carroll, S. (2021) *Something Deeply Hidden. Quantum Worlds and the Emergence of Spacetime*. London: Oneworld.

Conroy, Christina (2018) "Everettian actualism." *Studies in History and Philosophy of Modern Physics* 63, 24–33.

Craig, W. L. (2001). *Time and the Metaphysics of Relativity*. Dordrecht: Springer.

Crowther, K. (2016) *Effective Spacetime. Understanding Emergence in Effective Field Theory and Quantum Gravity*. Springer.

Dieks, D. (2012) "Time, Space, Spacetime." *Metascience* 21: 617–9.

Dyke, H. (2021) *Time*. Cambridge University Press.

Everett, H. (1957) ""Relative State" Formulation of Quantum Mechanics." *Review of Modern Physics* 29 (3): 454–62.

Ghirardi, G. C., Rimini, A., and Weber, T. (1986) "Unified dynamics for microscopic and macroscopic systems." *Physical Review D* 34, 440–91.

Glick, D. (2018) "Against Quantum Indeterminacy?" *Thought* 6(3), 204–213.

Greaves, H. (2004) "Understanding Deutsch's Probability in a Deterministic Multiverse." *Studies in History and Philosophy of Modern Physics* B35(3): 423–56.

Knox, E. (2013) "Effective spacetime geometry." *Studies in History and Philosophy of Modern Physics* 44, 346–56.

Le Bihan, B. (2020) "String theory, loop quantum gravity and eternalism." *European Journal for Philosophy of Science* 10: 1–22.

Lewis, D. K. (1986) *Philosophical Papers Vol. II*. New York: Oxford University Press.

Lucas, J. R. (1999) "A century of time." In Butterfield, J., editor, *The Arguments of Time*, 1–20. Oxford: Oxford University Press.

Mariani, C. and J. Torrengo (2021) "The indeterminate present and the open future." *Synthese* 199, 3923–3944.

Montague, R. (1962). "Deterministic theories." In N. F. Washburne (Ed.), *Decisions, values, and groups 2*, 325–370.

Peterson, D. and M. D. Silberstein (2010) "Relativity of Simultaneity and Eternalism." In Petkov (ed.), *Space, Time, and Spacetime*, 209–37. Heidelberg: Springer.

Plantinga, Alvin (1987) "Two concepts of modality: Moral realism and modal reductionism." *Philosophical Perspectives* 1, 189–231.

Putnam, Hilary (1967) "Time and Physical Geometry." *The Journal of Philosophy* 64 (8): 240–7.

Read, J. and E. Qureshi-Hurst (2021) "Getting tense about relativity." *Synthese* 198: 8103–25.

Rietdijk, C. W. (1966) "A Rigorous Proof of Determinism Derived from the Special Theory of Relativity". *Philosophy of Science* 33 (4): 341–4.

Romero, G. E. and D. Pérez (2014) "Presentism meets black holes." *European Journal for Philosophy of Science* 4: 293–308.

Rovelli, C. (2019) "Neither Presentism nor Eternalism." *Foundations of Physics* 49: 1325–35.

Saunders, S. (1993) "Decoherence, Relative States, and Evolutionary Adaptation." *Foundations of Physics* 23 (13), 1553–85.

Saunders, S. (1996) "Time, Quantum Mechanics, and Tense." *Synthese* 107, 19–53.

Saunders, S. (1998) "Time, Quantum Mechanics, and Probability." *Synthese* 114, 373–404.

Saunders, S. (2002) "How Relativity Contradicts Presentism." *Royal Institute of Philosophy Supplements* 50: 277–92.

Skyrms, B. (1980) *Causal Necessity*. New Haven: Yale University Press.

Slavov, M. (2020) "Eternalism and Perspectival Realism About the 'Now'". *Foundations of Physics* 50: 1398–1410.

Sudbery, A. (2017) "The logic of the future in quantum theory." *Synthese* 194: 4429–53.

Swinburne, R. (2008) "Cosmic simultaneity." In W. L. Craig and Q. Smith (eds.), *Einstein, relativity and absolute simultaneity*, 224–61. London: Routledge.

Tappenden, P. (2008) "Saunders and Wallace on Everett and Lewis." *The British Journal for the Philosophy of Science* 59 (3): 307–14.

Wallace, D. (2001) "Worlds in the Everett interpretation" arXiv:quant-ph/0103092v116 Mar 2001

Wilson, A. (2020) *The Nature of Contingency*. Oxford: Oxford University Press.
Zimmerman, Dean (2007) "The privileged present: Defending an 'a-theory' of time." In Sider, T. et al. (eds.), *Contemporary Debates in Metaphysics*, 211–25. Blackwell.